\documentclass[lettersize,journal]{IEEEtran}
\usepackage{amsmath,amsfonts}
\usepackage{algorithmicx}
\usepackage{algorithm}
\usepackage{array}
\usepackage[caption=false,font=normalsize,labelfont=sf,textfont=sf]{subfig}
\usepackage{textcomp}
\usepackage{stfloats}
\usepackage{url}
\usepackage{verbatim}
\usepackage{graphicx}
\usepackage{cite}
\usepackage{graphicx}
\usepackage{multirow}
\usepackage{amssymb}			

\usepackage{algorithm} 							
\usepackage{algpseudocode}				
\algrenewcommand\algorithmicrequire{\textbf{Input:}}
\algrenewcommand\algorithmicensure{\textbf{Output:}}

\begin{document}

\title{Robust Outlier Detection Method Based on Local Entropy and Global Density}

\author{Kaituo Zhang, Wei Huang, Bingyang Zhang, Jinshan Xu and Xuhua Yang
\thanks{This work was supported by the National Key R\&D Program of China (2022YFE0198900), the National Natural Science Foundation of China (61771430, 62176236), and the Natural Science Foundation of Zhejiang Province under grant LY22F020015.}
\thanks{Kaituo Zhang, Wei Huang, Bingyang Zhang, Jinshan Xu and Xuhua Yang are with the College of Computer Science, Zhejiang University of Technology, Hangzhou, 310023, China.}
\thanks{Wei Huang is the corresponding author (e-mail: huangwei@zjut.edu.cn).}}

\markboth{Zhang ET AL, Robust Outlier Detection Method Based on Local Entropy and Global Density}%
{Shell \MakeLowercase{\textit{et al.}}: A Sample Article Using IEEEtran.cls for IEEE Journals}


\maketitle

\begin{abstract}
By now, most outlier-detection algorithms struggle to accurately detect both point anomalies and cluster anomalies simultaneously. Furthermore, a few $K$-nearest-neighbor-based anomaly-detection methods exhibit excellent performance on many datasets, but their sensitivity to the value of $K$ is a critical issue that needs to be addressed. To address these challenges, we propose a novel robust anomaly detection method, called Entropy Density Ratio Outlier Detection (EDROD). This method incorporates the probability density of each sample as the global feature, and the local entropy around each sample as the local feature, to obtain a comprehensive indicator of abnormality for each sample, which is called Entropy Density Ratio (EDR) for short in this paper. By comparing several competing anomaly-detection methods on both synthetic and real-world datasets, it is found that the EDROD method can detect both point anomalies and cluster anomalies simultaneously with accurate performance. In addition, it is also found that the EDROD method exhibits strong robustness to the number of selected neighboring samples, the dimension of samples in the dataset, and the size of the dataset. Therefore, the proposed EDROD method can be applied to a variety of real-world datasets to detect anomalies with accurate and robust performances.
\end{abstract}


\section{Introduction}
\IEEEPARstart{O}{utlier} detection, also known as anomaly detection, initially emerged in the field of data mining and generally refers to the process of identifying a small number of anomalous samples within a set of normal samples. A substantial body of research has been devoted to defining data in the context of anomaly detection \cite{ref2}. In unsupervised learning, anomaly detection entails discovering patterns in data that deviate from expected behavior \cite{ref3,ref5}. In supervised learning, the task is to identify ways in which testing data differs from the data utilized during training \cite{ref4}. The detection of anomalous samples is of particular importance in various domains, as these samples can not only significantly impact data analysis but also greatly influence the accuracy of data predictions. As such, research on anomaly detection should concentrate on enhancing robustness, adaptability, and accuracy.

Anomalous samples are characterized by their infrequent occurrence and wide existence across various domains. Due to the diversity of data types, dimensions, and quantities in different fields, such as discrete sequences data \cite{ref23} and time series data \cite{ref42}, a universal anomaly detection method is elusive. Therefore, numerous studies have focused on the problem of anomaly detection in a wide range of real-world applications. For example, in intrusion detection, researchers have integrated anomaly detection algorithms into their studies \cite{ref7}, with a particular emphasis on areas such as the Internet of Things (IoT) and communication technologies \cite{ref8,ref9}. In image detection, specialized research focuses on applying anomaly detection algorithms to defect identification, examining textured surfaces like glass \cite{ref10}, fabric \cite{ref11}, and concrete \cite{ref12}, as well as detecting surface defects on printed circuit boards [13]. Anomaly detection has also been applied to medical image analysis \cite{ref14,ref15}. In fraud detection, studies have combined anomaly detection algorithms with e-commerce fraud detection \cite{ref16} and financial monitoring \cite{ref17}. As the amount of data continues to evolve with time, the scope of anomaly detection applications will undoubtedly keep expanding.


In recent years, numerous methods for anomaly detection have been proposed to address the increasing demand for anomaly detection in various applications. For example, the distribution-based methods \cite{ref4}, such as the Gaussian mixture model, aim to detect anomalies according to the distribution of data points. The graph-base method, such as benchmarking unsupervised outlier node detection (BOND) \cite{ref20}, out-of-distribution generalized graph neural network (OOD-GNN) \cite{ref18}, graph contrastive coding for anomaly detection (GCCAD) \cite{ref6} and augmentation for anomaly and normal distributions method (AugAn) \cite{ref22}, leverage graph structures to detect anomalies and outliers in the data. The distance-based methods, such as KNN and $K$-means algorithms, have the advantage of easy implementation and suitability for various types of data. The density-based methods, such as LOF, COF, KDE and DCROD \cite{ref25} algorithms, compare local densities of samples in the dataset to mine anomalies. Moreover, cluster purging \cite{ref1} is representative of rate-distortion theory-based outlier detection. Furthermore, deep learning-based methods have emerged as a major research direction, particularly with the use of deep autoencoders \cite{ref19,ref24} for anomaly detection.

Although the technology of anomaly detection has made substantial progress in recent years, there still exist some typical problems in current methods for anomaly detection. For example, in some existing methods, the performance of anomaly detection may be sensitive to the selection of parameter values. In other words, a slight deviation from the optimal parameter value may deteriorate the performance of anomaly detection. In KNN-based algorithms, the selection of values for the parameter $K$, i.e. the number of nearest neighbors, can lead to a significant effect on the performance of anomaly detection.

\begin{figure}
    \centering
    \includegraphics[width=0.6\linewidth]{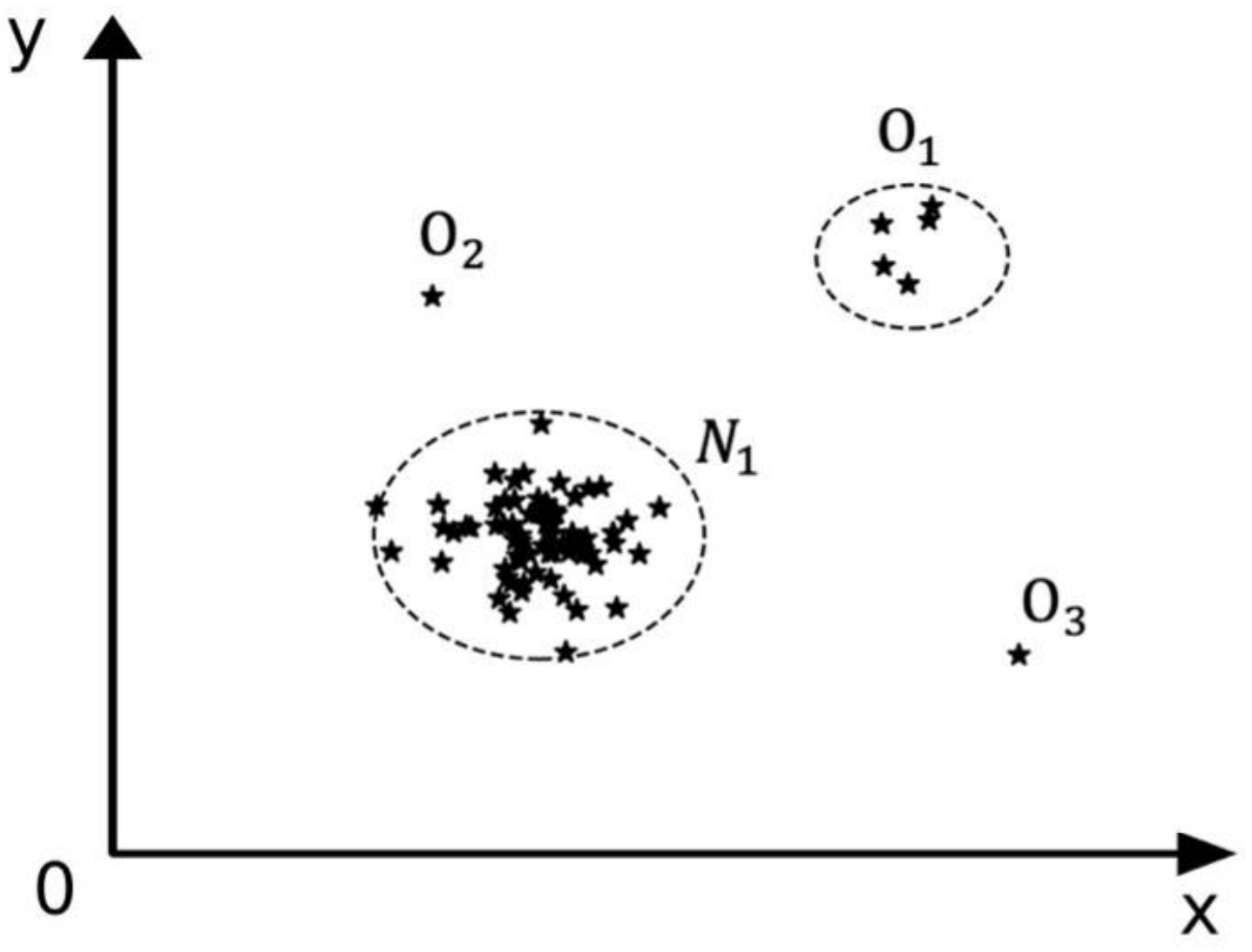}
    \caption{Illustration of point anomalies and cluster anomalies.}
    \label{fig:1}
\end{figure}

In addition, since point anomalies and cluster anomalies show different characteristics, it is not easy to simultaneously detect point anomalies and cluster anomalies. Here, pint anomalies are isolated points that reside outside the other majority of data points, while cluster anomalies occur when some discrete data points appear concentrated locally but are considered anomalous from a global perspective. Fig. \ref{fig:1} illustrates \(N_{1}\) as normal samples, \(O_{1}\) as a cluster anomaly, and \(O_{2}\) \& \(O_{3}\) as point anomalies.
Generally speaking, point anomalies are usually highly noticeable in the dataset, showing significant deviations in their local and global characteristics compared to normal samples. Consequently, traditional kernel density estimation (KDE) methods \cite{ref32} can be applied effectively for the detection of anomalies. However, detecting cluster anomalies presents more challenges. For example, KDE often struggles to identify the central sample in a cluster anomaly as an outlier. Hence, existing methods still encounter difficulties in simultaneously detecting point anomalies and cluster anomalies.

To address the above problems, we propose the Entropy Density Ratio Outlier Detection method, called EDROD method for short in this paper. This method incorporates the KDE method and the concept of Shannon entropy into the KNN configuration. The KDE method is used to extract global density information of each sample, while the Shannon entropy is used to depict the local information around each sample. By integrating global and local information of each sample into the task of anomaly detection, the proposed EDROD method has the following advantages:
\begin{itemize}
    \item \textbf{Effective detection of both point anomalies and cluster anomalies}: As we have analyzed above, in existing methods for anomaly detection, it is challenging to efficiently detect both point anomalies and cluster anomalies. The EDROD approach separately calculates the global density of the data and the local Shannon entropy, which holistically measures the local and global features of each data, thus effectively addressing the detection of both point anomalies and cluster anomalies.
    \item \textbf{Strong robustness to datasets and experimental conditions}: Through experiments on both synthetic and real-world datasets, it is found that the EDROD method exhibits strong robust performance in detecting anomalies. The enhanced robustness of EDROD is mainly reflected in the following three aspects. Firstly, $K$-nearest neighboring nodes based on Mahalanobis distance need to be selected for each sample. It is found drastic changes in $K$ value do not have an evident effect on the performance of anomaly detection, which indicates that EDROD is robust to the value of $K$. Secondly, the superiority of EDROD over several other competing algorithms is observed on both low-dimensional and high-dimensional datasets, which indicates that the EDROD method is also robust to the dimension of the dataset. Finally, the effectiveness of EDROD is verified on datasets with different sizes, which indicates that EDROD shows strong robustness to the size of datasets.
\end{itemize}

The rest of this paper is organized as follows. In Section II, we survey some methods for anomaly detection which are highly related to the proposed EDROD method. In Section III, we demonstrate the implementation process of the EDROD method in detail, followed by the analysis of computational complexities for the EDROD method. In Section IV, to validate the accuracy and robustness of the EDROD method, we compare the performances of EDROD method with those of several other competing algorithms for anomaly detection on both synthetic and real-world datasets. Finally, conclusions are drawn in Section V.

\section{Related Work}
In this section, we shall make a brief survey of some anomaly detection methods related to the proposed EDROD method. Since EDROD is a composite approach combing with distance-based, density-based, and information theory-based ideas, we shall only review these categories of methods for anomaly detection in this section.
\subsection{Distance-Based Anomaly Detection methods}
Distance-based anomaly detection methods are extensively employed techniques that focus on calculating distances among samples under different distance matrices to identify anomalies. The distance-based anomaly detection method utilized by EDROD is the widely renowned and popular KNN algorithm, which is widely embraced in the field of machine learning.

In the early stage after the KNN algorithm was proposed, the KNN algorithm primarily focused on density estimation and classification. When employing the KNN algorithm for these purposes, $K$-nearest neighbors of each sample are required to be selected to implement the task of density estimation and sample classification. Scholars have utilized various distance metrics, such as Euclidean distance and Mahalanobis distance, to adapt to different data types for classification \cite{ref33}. The choice of the $K$ value has a significant impact on the performance of the KNN algorithm. A large value of $K$ increases computational costs and affects the performances of the KNN algorithm \cite{ref26}, while a small value of $K$ may not adequately represent sample characteristics and capture relationships among samples.

Later, the KNN algorithm was utilized for anomaly detection. Initially, the KNN algorithm was used to compute weights for each sample based on the total distances of this sample to its $K$-nearest neighbors. Then, outliers can be picked out from all samples according to the weights of samples \cite{ref27}. However, relying solely on the KNN algorithm for anomaly detection has proven to be of low robustness and limited adaptability. Recent algorithms often employ the KNN algorithm as an intermediate step of other tasks such as density calculation \cite{ref25,ref28} and the angle-based outlier detection (ABOD) algorithm \cite{ref39}.

\subsection{Density-Based Anomaly Detection methods}
The density-based anomaly-detection method, which has been popularly used to detect anomalies in real-world systems, usually incorporates some other types of anomaly-detection methods, such as probability-based method and distance-based method. The EDROD method we propose in this paper also incorporates the process of density estimation, which is used to capture the global feature of each sample to help determine anomalies.

The capability of the KNN algorithm to compute the local features of samples has made it extensively utilized in density-based anomaly detection methods. The Local Outlier Factor (LOF) algorithm, which was proposed in \cite{ref29} and applies the concept of the KNN algorithm, is a classical density-based method of anomaly detection. The LOF algorithm assigns a local outlier factor (LOF) to each sample in the dataset, indicating the degree of outlier for each sample. However, the LOF algorithm has the drawback that the LOF algorithm solely considers the difference between the density of each sample and those of its surrounding neighbors. Thus, the effectiveness of the LOF algorithm will be degraded when the density of one sample is close to those of its neighbors. To address this problem, \cite{ref30} proposed the connectively-based outlier factor (COF) algorithm, which calculates local density based on the so-called average chain distance. However, the improvement of the COF algorithm over the LOF algorithm is with the sacrifice of increased computational complexity. To address challenges associated with extensive memory usage and poor performance of the LOF algorithm on high-dimensional datasets, \cite{ref31} introduced the combined entropy and LOF (CELOF) algorithm which improves the LOF algorithm by incorporating the concept of entropy in information theory. Empirical studies in \cite{ref31} show that the CELOF approach significantly enhances the anomaly detection performance of the LOF method on high-dimensional datasets.

In contrast to LOF, COF, and CELOF which compare the local density of data points with their neighbors to determine anomalies, the Kernel Density Estimation (KDE) method directly estimates the density of each data sample to determine anomalies. The underlying idea of the KDE method is that outliers typically exhibit distinctive "peaks" or "valleys" in the probability density function compared to normal data. Later on, combing the LOF algorithm with the KDE algorithm, \cite{ref32} proposed the Local Density Factor (LDF) method, which utilizes KDE for local density estimation and calculates distances among local samples to obtain an evaluation metric for anomaly detection. In addition, \cite{ref37} proposed the robust kernel density estimation (RKDE) method by combining the traditional KDE algorithm with the idea of M-estimation. Both theoretical analysis and experimental results demonstrate the robustness of the RKDE method to outliers.

\subsection{Information Theory-Based Anomaly Detection methods}
The basic idea of anomaly detection based on information theory is to formulate the problem of anomaly detection into an optimization problem, seeking to identify a set of anomalies that maximize or minimize information gain. Typically, the information entropy is usually employed as a measure of information gain or uncertainty in these methods \cite{ref34}.

Leveraging the advantages of entropy calculation for both discrete and continuous data, \cite{ref35} introduced the Local Search Algorithm (LSA) method. This method generates combinations of samples by considering one sample as the center, forming local feature samples. Based on the probabilities of samples, the information entropy of each combination of samples can be calculated. Then, the one with the minimum information entropy is selected, and the center sample is identified as the anomaly. However, constructing all possible sample combinations in LSA results in high computational complexity, especially on large datasets. To address the problem of high computational complexities, \cite{ref36} proposed the fast greedy method to detect anomalies. Experimental results on both large synthetic and real-world datasets show that, as compared to the LSA method, the greedy algorithm can detect anomalies at a much faster speed while still achieving comparable performances in the precision of anomaly detection.

\section{The Proposed Method}
In this section, we propose a highly robust method, denoted by Entropy Density Ratio Outlier Detection (EDROD) method, to perform anomaly detection in effective way. This method calculates the anomaly score for each data sample by incorporating two measurements. The first measurement is the sample density calculated using the technique of Kernel Density Estimation (KDE). Then, based on sample densities, local entropy is calculated around each sample to measure the distribution of local densities. For each sample, the ratio between local entropy and density (Entropy Density Ratio, \textit{EDR}) is calculated as the anomaly score to assess the degree of abnormality.
\begin{figure*}[!t]
\centering
{\includegraphics[width=7in]{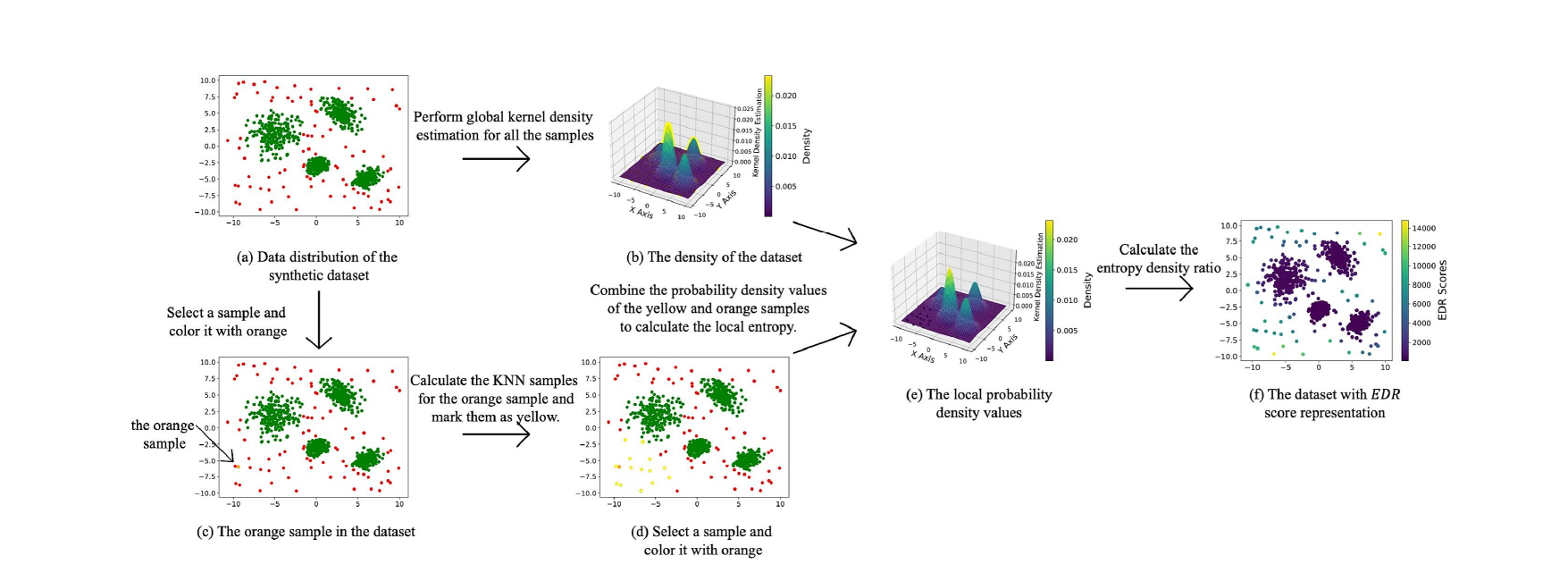}%
\label{fig_second_case}}
\caption{Structure of the proposed EDROD method in the two-dimensional case.}
\label{fig:2}
\end{figure*}

Fig. \ref{fig:2} illustrates the implementation process of the EDROD method using several artificial-generated data points on the two-dimensional plane as an example. As shown in Fig. \ref{fig:2}(a), normal and abnormal samples are marked with green and red respectively. Then, using the technique of kernel density estimation, the densities of all samples are calculated, as shown in Fig. \ref{fig:2}(b). It can be clearly observed from Fig. \ref{fig:2}(b) that the high peaks in the three-dimensional space correspond to high-density regions in the two-dimensional plane as shown in Fig. \ref{fig:2}(a). To show the process of calculating local entropy, we randomly select a specific sample \textit{p} and marked it with orange as illustrated in Fig. \ref{fig:2}(c). Then, based on the metric of Mahalanobis distance, we select 14 nearest neighboring samples of the sample \textit{p}. The 14 nearest neighboring samples are colored with yellow as illustrated in Fig. \ref{fig:2}(d). Then, the local entropy of the sample $p$ is calculated based on the densities of sample \textit{p} and its 14 nearest neighboring samples. Fig. \ref{fig:2}(b) shows the densities of samples and Fig. \ref{fig:2}(e) shows the local probability density values of local samples. Using the local probability density, we calculate the local entropy, and we finally calculate the \textit{EDR} score by dividing the local entropy of each sample by its global density. The brighter color indicates a higher \textit{EDR} score and a higher likelihood of being an anomaly.

In the following, we shall describe the details of each step for our proposed EDROD method.

\subsection{Global kernel density estimation}
The task of anomaly detection mainly aims to identify anomalous samples in the dataset. These samples are usually far away from clustered samples and isolated in areas with fewer samples. By using the technique of density estimation, the relatively low-density regions where anomalous samples are located can be detected from the dataset. Due to the high estimation precision and easy implementation, the KDE method is utilized to perform global density estimation for the given dataset.

KDE is a non-parametric technique to estimate the probability density of each sample in a group of samples. The general idea of KDE is to place a kernel function around the data sample. Then, kernel functions of all samples are accumulated to yield the final probability density estimation of each sample. Assuming there are totally $n$ $d$-dimensional samples in the dataset, the expression of probability density $\hat{\rho}(\textbf{x})$ for a single sample \textbf{x}, is shown as:
\begin{equation}
\label{eq1}
\hat{\rho}(\mathbf{x}) = \hat{\rho}(x_1, \ldots, x_d) = \frac{1}{{nh^d}} \sum_{i=1}^{n} \Phi\left(\frac{\mathbf{x}-\mathbf{x}_i}{h}\right),
\end{equation}
where $h$ represents the preset kernel width of the kernel function and $\Phi(\cdot)$ represents the kernel function. The most widely used kernel functions include the Gaussian function, radial basis function, linear kernel function \textit{etc}. Due to the excellent performance of fitting ability and generalization ability, the Gaussian kernel function is used in the EDROD method to estimate sample densities. The expression of the Gaussian kernel function is shown as
\begin{equation}
\label{eq2}
\Phi(\textbf{x}) = \frac{1}{{(2\pi)^d}} \exp\left(\frac{-\|\textbf{x}\|^2}{2}\right),
\end{equation}
where $\|\textbf{x}\|$ represents the $l_2$-norm of vector \textbf{x}.

With \eqref{eq1} and \eqref{eq2}, we have the following expression for the probability density of sample $\mathbf{x}_i$:
\begin{equation}
\label{eq3}
\hat{\rho}(\textbf{x}_i) = \frac{1}{{nh^d}} \sum_{j=1, j\neq i}^{n} \frac{1}{{(2\pi)^d}} \exp\left(-\frac{\|\textbf{x}_i-\textbf{x}_j\|^2}{2h^2}\right)
\end{equation}

\begin{figure}[ht]
    \centering
    \includegraphics[width=3in]{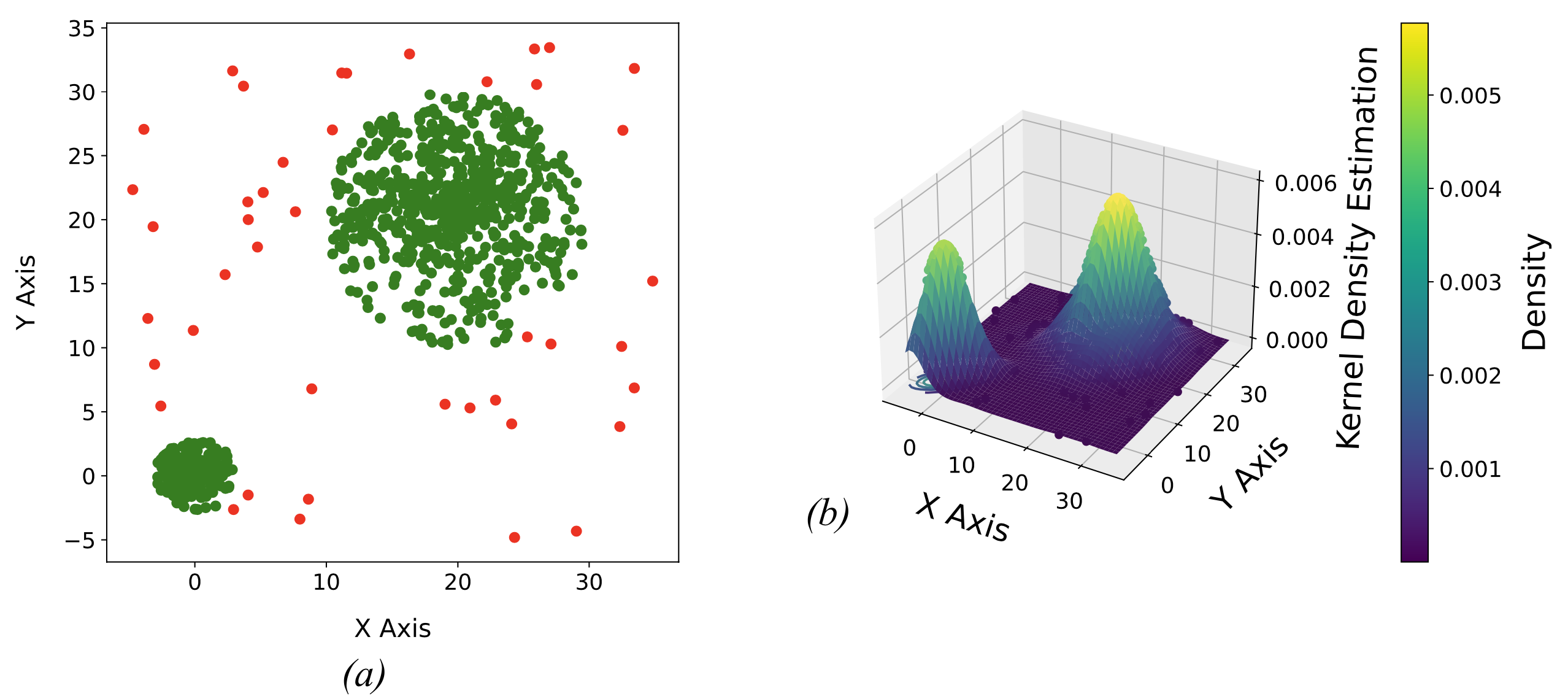}
    \caption{A toy example of kernel density estimation.}
    \label{fig:3}
\end{figure}

Fig. \ref{fig:3} shows a toy example of kernel density estimation for the two-dimensional dataset. Fig. \ref{fig:3}(a) is the scatter plot of some data points on the two-dimensional plane and Fig. \ref{fig:3}(b) illustrates the densities of these points in the three-dimensional space. It can be clearly observed that areas with dense samples in Fig. \ref{fig:3}(a) correspond to regions with high-density estimates in Fig. \ref{fig:3}(b), while areas with sparse samples in Fig. \ref{fig:3}(a) correspond to regions with low-density values in Fig. \ref{fig:3}(b).

\subsection{Selection for $K$-nearest neighbors based on Mahalanobis distance}
Next, we shall calculate the local entropy of each sample based on the densities of samples in the local region. Thus, we need to choose $K$-nearest neighbors for each sample. It should be noted that the selection of distance metrics is important in the KNN algorithm since the set of $K$ nearest neighbors may not be totally the same under different distance metrics. Since real-world data usually have multiple dimensions and distributions of data points usually vary from each on different directions, we use Mahalanobis distance rather than Euclidean distance to select the $K$-nearest neighbors of each sample. Compared to the Euclidean distance, Mahalanobis distance takes into account the correlation among various directions. Given two samples, denoted as $\mathbf{x}_i$ and $\mathbf{x}_j$, the Mahalanobis distance $Md$($\mathbf{x}_i$, $\mathbf{x}_j$) between $\mathbf{x}_i$ and $\mathbf{x}_j$ is shown as
\begin{equation}
\label{eq4}
Md(\textbf{x}_i, \textbf{x}_j) = \sqrt{ (\textbf{x}_i - \textbf{x}_j)^T \Sigma^{-1} (\textbf{x}_i - \textbf{x}_j) },
\end{equation}
where $\Sigma$ denotes the covariance matrix of all data points in the dataset.

With the metric of Mahalanobis distance, we now improve the traditional KNN algorithm which selects the $K$-nearest neighbors based on Euclidean distance. For the sample $\mathbf{x}_i$, we sort the Mahalanobis distances between sample $\mathbf{x}_i$ and other samples in the dataset, and select $K$-nearest samples as the neighbors of sample $\mathbf{x}_i$. Given the dataset $X=\{\textbf{x}_1, \textbf{x}_2, \textbf{x}_3, \ldots, \textbf{x}_n\}$, where $\textbf{x}_i \in \mathbb{R}^d$ ($i$=1, 2, 3, $\ldots$, $n$) and $d$ is the dimension of the data, we define the expression of Md-KNN (Mahalanobis distance-based KNN) is shown as
\begin{equation}
\label{eq5}
Knn(\textbf{x}_i, K) = S(Md(\textbf{x}_i, \textbf{x}_j), K), \quad \textbf{x}_j \in X, \quad \textbf{x}_j \neq \textbf{x}_i,
\end{equation}
where $Knn(\textbf{x}_i, K)$ represents the set of $K$-nearest neighbor samples of $\textbf{x}_i$. Additionally, $S(Md(\textbf{x}_i, \textbf{x}_j), K)$ is a function used to select the $K$-nearest samples based on Mahalanobis distance. Here, $\textbf{x}_j$ represents another sample in the dataset which is different from $\textbf{x}_i$.

Selecting the $K$-nearest neighbors of each sample can be used to extract the local features around each sample. This operation links the features of each sample with those of its nearby samples, and enables the incorporation of local information into the feature representation of each sample.

\subsection{Local Entropy}
While considering the influence of anomalous samples within the global dataset, it remains essential to consider the relationships between individual samples and their local neighbors in the task of anomaly detection. This is mainly due to the potential occurrence of cluster anomalies, which further necessitates more comprehensive analysis.

By using the Md-KNN approach, we create a local group consisting of each sample and its $K$-nearest neighbors. After conducting kernel density estimation for all samples in the whole dataset, every sample within the group is endowed with a global probability density. Consequently, by computing the Shannon entropy based on the global probability density of all samples in this group, we are able to effectively portray the local characteristics of the present group around one central sample.

In the realm of continuous probability density, obtaining probability values necessitates the use of integration. However, as practical datasets predominantly consist of discrete data, conventional methods are not applicable for acquiring probability values. Instead, we opt to normalize the probability density values of samples situated within a group centered around $\textbf{x}_j$ and obtain the normalized probability densities of all samples in this group, which is demonstrated by the following expression:
\begin{equation}
\label{eq6}
\tilde{\rho}(\textbf{x}_{ij}) = \frac{\hat{\rho}(\textbf{x}_{ij})}{\sum_{l=1}^{K} \hat{\rho}(\textbf{x}_{il}) + \hat{\rho}(\textbf{x}_{ii})},
\end{equation}
where $\textbf{x}_i, \textbf{x}_l \in Knn(\textbf{x}_i, K) \cup \textbf{x}_i$, $\textbf{x}_{ij}$ denotes the $j$-th nearest neighbor of $\textbf{x}_i$, and $\tilde{\rho}(\textbf{x}_{ij})$ represents the normalized probability density of $\textbf{x}_{ij}$ within the group around the central sample $\textbf{x}_i$.

After obtaining the normalized probability density within the local group, values of all probability densities fall within the range between 0 and 1. It is easy to see that all normalized probability densities in the same local group sum to one. Therefore, we can use the concept of Shannon entropy to measure the heterogeneity of normalized probability in one local group around the central sample $\textbf{x}_i$, which can be expressed as:
\begin{equation}
\label{eq7}
E(\textbf{x}_i) = -\tilde{\rho}(\textbf{x}_{ii}) \log(\tilde{\rho}(\textbf{x}_{ii})) - \sum_{j=1}^{K} \tilde{\rho}(\textbf{x}_{ij}) \log(\tilde{\rho}(\textbf{x}_{ij})).
\end{equation}

Based on the definition of Shannon entropy, more uniform probability distribution indicates a closer likelihood of various events, which in turn increases uncertainty and leads to a larger entropy. On the other hand, a more concentrated probability distribution, where the occurrence probability of some events is significantly higher than others, reduces uncertainty and results in a smaller value of entropy. Consequently, when local samples are densely packed, their probabilities are closer, yielding a larger value of $E(\textbf{x})$. In contrast, when outliers are present in local samples and probability density is normalized, a concentrated probability distribution emerges, leading to a smaller value of $E(\textbf{x})$.

By computing entropy, we establish a new cue for detecting cluster anomalies. Specifically speaking, when such anomalies arise, the samples become less concentrated, leading to an increase in Shannon entropy. After computing the entropy, we have the information of local sample characteristics that allow us to differentiate between outlier and clustered samples. If the sample is an outlier, the local entropy around this sample is relatively small. In contrast, if the sample exhibits a clustering feature, the local entropy around this sample is relatively large.

\subsection{Ratio between Shannon entropy and density}
As both normal samples and cluster anomalies can exhibit clustering structures, it is sometimes difficult to distinguish cluster anomalies from normal samples. Thus, it remains crucial to consider the features of samples within the global dataset.

We propose that by using both the entropy and probability density of samples, we can efficiently harness the local and global characteristics of samples in the dataset. Consequently, we put forward the concept of Entropy Density Ratio (\textit{EDR}) for each sample $\textbf{x}_i$, which is expressed as:
\begin{equation}
\label{eq8}
EDR(\textbf{x}_i) = \frac{E(\textbf{x}_i)}{\hat{\rho}(\textbf{x}_i)},
\end{equation}
where $E(\textbf{x}_i)$ is the local entropy of $\textbf{x}_i$ and $\hat{\rho}(\textbf{x}_i)$ is the probability density of $\textbf{x}_i$, which is defined in \eqref{eq1} using the technique of kernel density estimation.

The measurement $EDR(\textbf{x}_i)$ effectively considers both global and local features of sample $\textbf{x}_i$, making it highly proficient in detecting both outlier anomalies and cluster anomalies. In the case of outlier anomalies, the probability density is smaller, leading to a larger $EDR(\textbf{x}_i)$ compared to normal samples. When sample $\textbf{x}_i$ exhibits the cluster anomaly, the local entropy of sample $\textbf{x}_i$ is larger, but its probability density is smaller than those of normal samples, resulting in a larger $EDR(\textbf{x}_i)$ compared to normal samples. Finally, since both probability densities and local entropies of normal samples are relatively higher, normal samples usually cannot result in larger $EDR(\textbf{x}_i)$ than cluster anomalies and point anomalies.

In our proposed EDROD method, we use \textit{EDR} as the anomaly score to determine whether a testing sample is abnormal or not. Since both local and global characteristics of all samples may vary from each other, as has been analyzed above, it is easy to understand that if the sample $\textbf{x}_i$ is abnormal, its \textit{EDR} score will be larger. Therefore, we can sort \textit{EDR} scores of samples in descending order and consider samples with larger scores as anomalous samples.

The overall pseudocode of the EDROD algorithm is shown in Algorithm 1.

\begin{algorithm}[h]
    \caption{Entropy Density Ratio Outlier Detection (EDROD)}
    \begin{algorithmic}[1]
    \Require Dataset $\mathbf{X}$, the size $K$ of nearest neighbors
    \Ensure Outlier scores \textit{EDR}
    \State Initialization $EDR \leftarrow \{\}$;
    \For{\textit{every} $\mathbf{x}_i \in \mathbf{X}$}
        \State Obtain global density $\hat{\rho}(\mathbf{x}_i)$ using \eqref{eq3};
    \EndFor
    \For{\textit{every} $\mathbf{x}_i \in \mathbf{X}$}
        \For{every $\mathbf{x}_j \in \mathbf{X}$ and $\mathbf{x}_j \neq \mathbf{x}_i$ }
            \State Obtain $Md(\mathbf{x}_i, \mathbf{x}_j)$ of $\mathbf{x}_i$ and $\mathbf{x}_j$ using \\ \quad \quad \quad Mahalanobis distance;
        \EndFor
        \State $Knn(\mathbf{x}_i, K) = S(Md(\mathbf{x}_i, \mathbf{x}_j), K)$;
    \EndFor
    \For{\textit{every} $\mathbf{x}_i \in \mathbf{X}$}
        \For{every $\mathbf{x}_{ij} \text{ in } Knn(\mathbf{x}_i, K)$}
            \State Calculate $\tilde{\rho}(\mathbf{x}_{ij})$ using \eqref{eq6};
        \EndFor
        \State Calculate local entropy $E(\mathbf{x}_i)$ of $\textbf{x}_i$ using \eqref{eq7};
    \EndFor
    \For{\textit{every} $\mathbf{x}_i \in \mathbf{X}$}
        \State Calculate $EDR(\textbf{x}_i)$ through $\hat{\rho}(\mathbf{x}_i)$ and $E(\mathbf{x}_i)$ using \eqref{eq8};
    \EndFor
    \State Output the outlier scores \textit{EDR}.
    \end{algorithmic}
    \label{alg:syn}
\end{algorithm}

\begin{figure}
    \centering
    \includegraphics[width=2.5in]{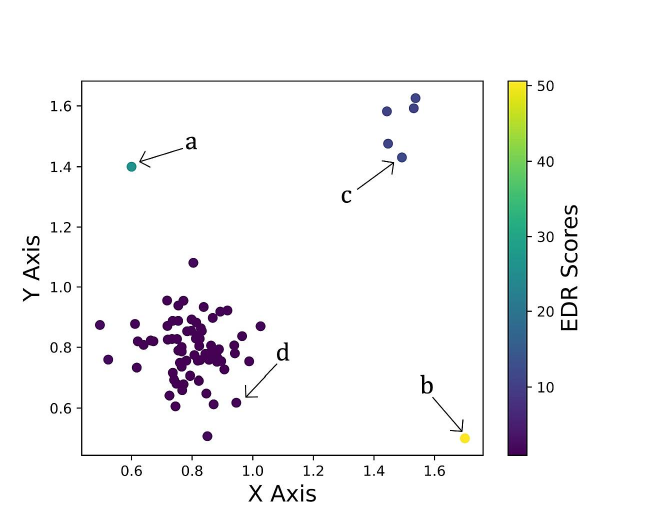}
    \caption{Heatmap of \textit{EDR} score for data points on the two-dimensional plane.}
    \label{fig:4}
\end{figure}

To demonstrate the role of \textit{EDR} score in distinguishing anomalous samples from normal samples more clearly, we apply the EDROD method to a toy example with several data points on the two-dimensional plane. We calculate \textit{EDR} scores for all samples and visualize \textit{EDR} scores on a heatmap which is shown in Fig. \ref{fig:4}. In this heatmap, darker colors indicate lower \textit{EDR} scores, which correspond to normal samples. Lighter colors indicate higher \textit{EDR} scores, which correspond to anomalous samples.

In Fig. \ref{fig:4}, we select four typical samples to illustrate the effectiveness of the EDROD method, among which samples \textit{a}, \textit{b}, \textit{c}, and \textit{d} are point anomaly, point anomaly, cluster anomaly, and normal sample respectively. We start our analysis from samples \textit{a} and \textit{b}. As both samples \textit{a} and \textit{b} are located far away from the clustered samples, their probability density values are low. In addition, both samples \textit{a} and \textit{b} exhibit significant differences in probability density from those of their nearest neighboring samples. Consequently, the local Shannon entropies based on normalized probability densities around both sample \textit{a} and sample \textit{b} are smaller. However, when considering \textit{EDR} score calculation, we also need to take into account the global probability density. Consequently, according to \eqref{eq8}, a high Shannon entropy over a small probability density results in larger \textit{EDR} scores. Since sample \textit{b} is further away from the other samples compared to sample \textit{a}, it is more likely to be an anomaly, resulting in a higher \textit{EDR} score for sample \textit{b}. Now let us consider sample \textit{c}. This sample represents the type of cluster anomaly, which is formed by a cluster of partially anomalous points. When examined locally, sample \textit{c} appears to be a normal sample because the probability densities of its neighboring points are similar to that of sample \textit{c}, resulting in relatively large local entropy. However, from the global perspective, the probability density of sample \textit{c} is lower than those of normal samples. Consequently, with high local entropy and relatively low probability density, the sample \textit{c} has a high \textit{EDR} score. Finally, we move on to sample \textit{d}. It can be analyzed that the sample \textit{d} has a relatively low \textit{EDR} score. This is because, although the probability density values of its $K$-nearest neighbors are close to each other, leading to higher local entropy, the probability densities of sample \textit{d} and its $K$-nearest neighbors are larger. As a result, the \textit{EDR} score of sample \textit{d} is relatively small.

The introduction of local Shannon entropy allows for the utilization of the local characteristics of samples, which helps address the issue of poor robustness in KNN-based methods due to changes in the parameter $K$. EDROD also incorporates the global characteristics of samples, i.e. the global probability densities, which enables the detection process to transcend local features and be more effective in addressing cluster anomaly scenarios.

\subsection{Complexity analysis}
In this section, we shall analyze the computational complexity of the EDROD method according to the implementation steps of 4. Denoting the number of samples in the dataset as $n$, the number of features on each sample as $d$, and the number of nearest neighboring samples as $K$, computational complexities at each step of EDROD are summarized as follows:
\begin{enumerate}
    \item Calculating Mahalanobis distance: Since Mahalanobis distances need to be calculated among each pair of samples, the computational complexity at this step is around $O(n^2)$.
    \item Global kernel density estimation: The computational complexity of global kernel density estimation mainly depends on the size of the dataset and the number of features. Therefore, the computational complexity of global kernel density estimation is around $O(nd)$.
    \item Local entropy: For each sample, the local entropy is calculated based on the densities of this sample and its $K$-nearest neighboring neighbors. Therefore, the time complexity of calculating local entropy is $O(nK)$.
\end{enumerate}

 Therefore, since $n$ is usually much larger than both $d$ and $K$, the computational complexity of EDROD can be approximated as $O(n^2 d) + O(nd) + O(nK) \approx O(n^2 d)$.

\section{Experiment results and relevant discussions}
In this section, we will apply the proposed EDROD algorithm to several datasets and evaluate its performance in the task of anomaly detection. We shall compare the performances of the EDROD method with those of some other competing algorithms to comprehensively assess the effectiveness and advantages of the EDROD method.

\subsection{The compared methods}
The proposed EDROD method firstly aims to address the problem of inefficiency and weak robustness of KNN-based methods in detecting anomalies on high-dimensional datasets. Therefore, we shall first compare the EDROD method with other KNN-based methods. In our experiments, we shall select KNN \cite{ref38}, LOF \cite{ref29}, COF \cite{ref30}, ABOD \cite{ref39}, and DCROD \cite{ref25} methods to compare their performances with that of the EDROD method.

Besides, we shall also compare the EDROD method with other categories of methods, which include OCSVM \cite{ref40}, COPOD \cite{ref41}, KDE \cite{ref32}, PCA \cite{ref43}, IForest \cite{ref44}, ECOD \cite{ref45}, and LUNAR \cite{ref46} methods. Among these, OCSVM, KDE, and PCA are early classical methods, while COPOD, ECOD, and LUNAR methods are more recent approaches. Both COPOD and ECOD methods consider the existence of correlations among high-dimensional data and employ empirical correlation methods, whereas the LUNAR method utilizes a graph neural network for anomaly detection. We use these methods to validate the performance of the EDROD method in detecting anomalies. In our experiments, we set the kernel width in KDE to 1.0, the number of base estimators in IForest to 100, and the learning rate in LUNAR to 0.001.

\subsection{Datasets}
In our study, we carry out experiments on both synthetic and real-world datasets. In the synthetic datasets, we consider both two-dimensional and ten-dimensional cases. The two-dimensional dataset, which consists of 712 normal samples and 130 anomalous samples, is used to visually demonstrate the capability of the EDROD method in simultaneously detecting both outlier and cluster anomalies.

In the 10-dimensional synthetic dataset \cite{ref47}, there are two types of data with different sizes of datasets. The first type consists of 270 normal samples and 30 anomalous samples (called \textit{300\_SAMPLE} dataset for short in this paper), while the second type comprises 630 normal samples and 70 anomalous samples (called \textit{700\_SAMPLE} dataset for short in this paper). In each type of data, we test the performance of all algorithms under study under 10 independent dataset instances. Each result is the average over the 10 independent dataset instances.

Besides, to validate the real-world applicability and reliability of the EDROD algorithm, we have also conducted studies on a set of real-world datasets for all methods under study. We select nine real-world datasets that exhibit high dimensionality, irregular distribution, and varying proportions of anomalous samples. The specific information of these datasets is presented in TABLE \ref{tab:1}. These datasets are available in our github\footnote{https://github.com/Philip0512/EDROD}.

\begin{table}[ht]
\centering
\caption{Information about nine real-world datasets}
\label{tab:1}
\begin{tabular}{l|cccc}
\hline
\textbf{Dataset} & \textbf{Dimension} & \textbf{Samples} & \textbf{Outliers} & \textbf{Outlier Percentage} \\ \hline
Wind Turbine & 12 & 942 & 7 & 0.7\% \\
Wave & 40 & 742 & 12 & 1.6\% \\
Breastw & 9 & 683 & 239 & 34.9\% \\
WDBC & 30 & 350 & 50 & 14.3\% \\
Wine & 13 & 129 & 10 & 7.8\% \\
Letter & 32 & 1600 & 100 & 6.3\% \\
ionosphere & 33 & 351 & 126 & 35.9\% \\
Glass & 9 & 214 & 9 & 4.2\% \\
Parkinson & 22 & 195 & 147 & 75.4\% \\ \hline
\end{tabular}
\end{table}

\subsection{Evaluation measurement}

In our study, the AUC (Area Under the Curve) value serves as an indicator of how well an algorithm can discriminate between normal and abnormal samples. By comparing the AUC values of all competing algorithms, we can evaluate their relative performances with each other. In addition, by observing how the AUC value changes with varying the value of $K$ for all KNN-based algorithms, we can gain insights into the robustness and stability of these algorithms under different conditions. Finally, the performances of KNN-based algorithms are evaluated by averaging AUC values with the parameter \textit{K} in some intervals to ensure a fair and comprehensive comparison.

\subsection{Experimental results and discussions on synthetic datasets}
\begin{figure}
    \centering
    \includegraphics[width=0.6\linewidth]{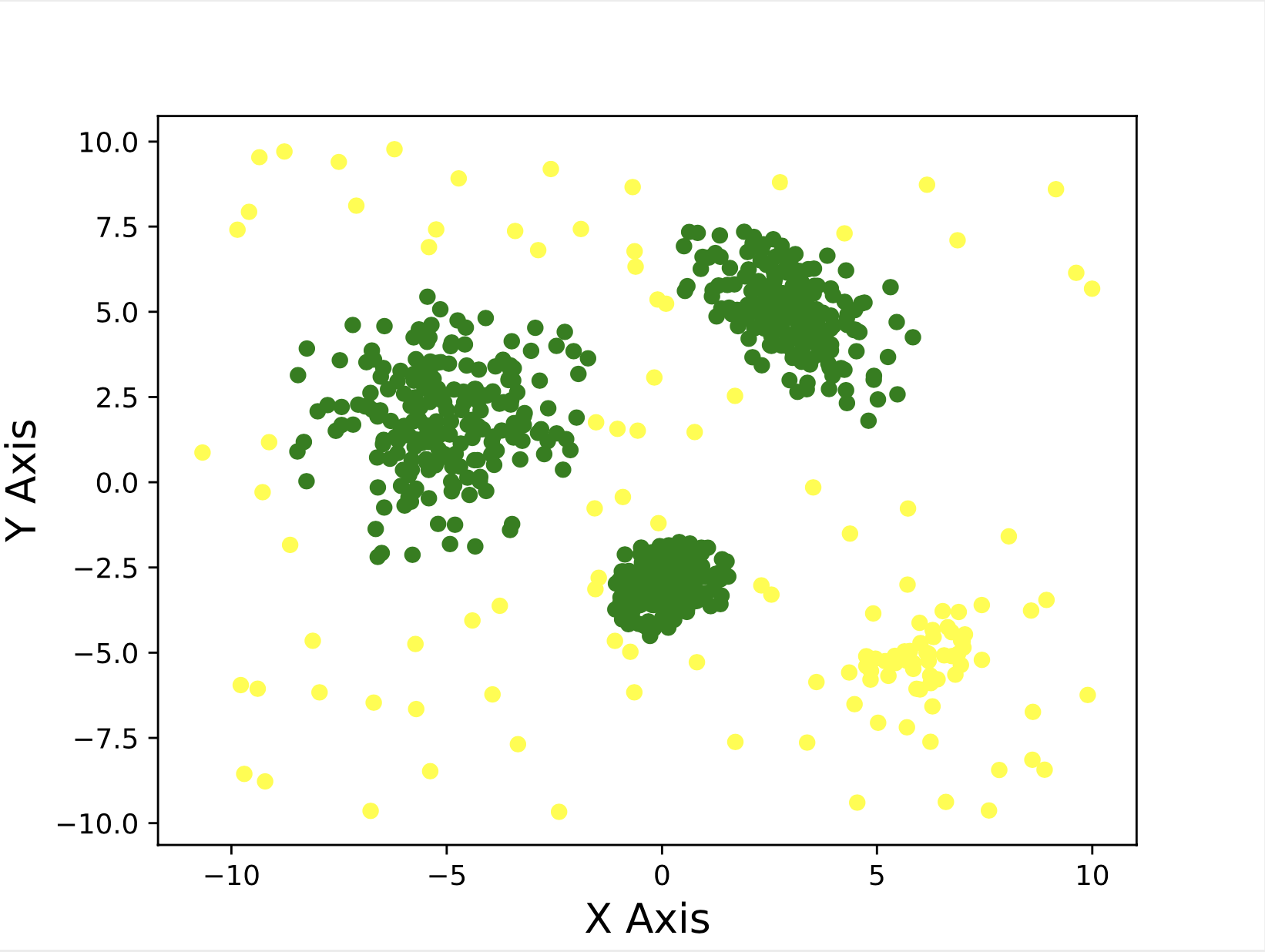}
    \caption{Visualization of the two-dimensional dataset.}
    \label{fig:5}
\end{figure}

As we have stated in the section of the introduction, our proposed method has the remarkable prowess of detecting both point anomalies and cluster anomalies simultaneously. To demonstrate this advantage in a more intuitive way, we first compare the performances of anomaly detection for the EDROD method and its competing counterparts on the two-dimensional dataset. We handpick the meticulously curate two-dimensional dataset with 712 normal samples and 130 anomalous samples. These samples are gracefully arranged in the two-dimensional plane, which is visualized in Fig. \ref{fig:5}. Normal and anomalous samples are colored with green and red respectively. Notably, within this remarkable dataset, we can find a series of point anomalies (colored in red), which are scattered in the two-dimensional plane. Besides, on this dataset, there exist a collection of cluster anomalies, which is nestled in the lower-right region of the two- dimensional plane.

On this dataset, we carry out the task of anomaly detection using all methods mentioned above. Values of parameter $K$ in all methods studied here are set to 20 to yield the optimal performance using the grid search method. We calculate AUC values for all methods, and the results are demonstrated in TABLE \ref{tab:2}. It can be observed that our proposed EDROD method outperforms all other competing methods, showing the superior effectiveness of the EDROD method in anomaly detection. The KDE, KNN and IForest methods yield AUC values slightly smaller than that of the EDROD method. However, AUC values of the remaining algorithms are much smaller than that of the EDROD method.
\begin{table}[ht]
\centering
\caption{AUC values of the EDROD method and its competing counterparts on the two-dimensional dataset}
\label{tab:2}
\begin{tabular}{lc|lc}
\hline
\textbf{Approach} & \textbf{AUC Values} & \textbf{Approach} & \textbf{AUC Values} \\ \hline
KNN & 0.942 & KDE & 0.954 \\
LOF & 0.777 & PCA & 0.861 \\
COF & 0.765 & IForest & 0.956 \\
ABOD & 0.919 & ECOD & 0.897 \\
DCROD & 0.914 & LUNAR & 0.898 \\
OCSVM & 0.928 & \textbf{EDROD} & \textbf{0.959} \\
COPOD & 0.785 & & \\
\hline
\end{tabular}
\end{table}

\begin{figure*}[!t]
\centering
    \includegraphics[width=1\linewidth]{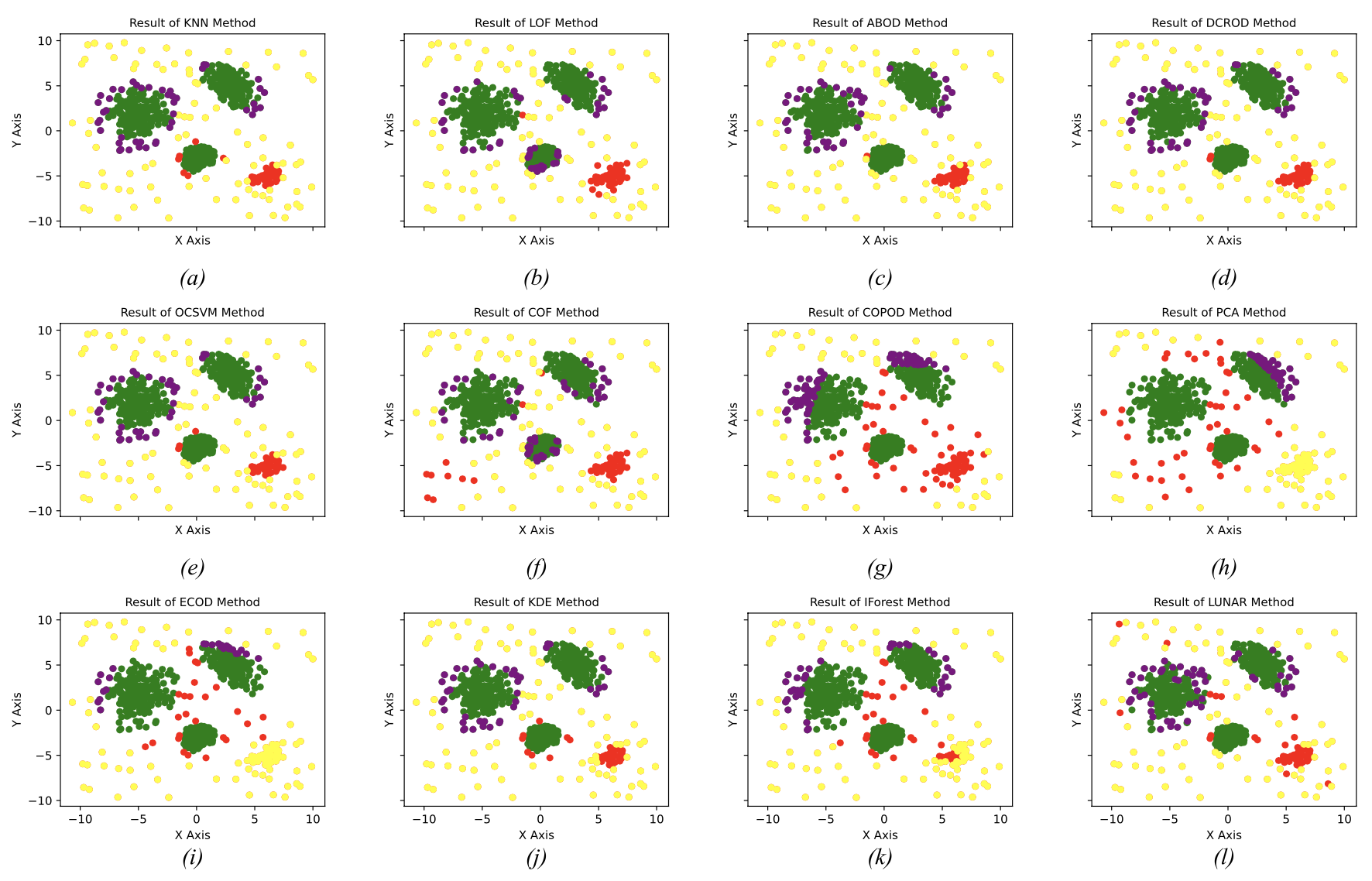}
\caption{Visualization of detection results for the competing counterparts on the two-dimensional dataset.}
\label{fig_6}
\end{figure*}

To facilitate us comparing the effectiveness of all methods in detecting anomalies more thoroughly, we select 130 samples that are identified as anomalies by various methods, which is rightly equal to the number of anomalous samples in the original dataset. Results of all methods are demonstrated in Fig. \ref{fig_6}. For each method, the samples are marked with different colors according to the combination of “original sample category” and “detection result”, which are shown in TABLE \ref{tab:3}.

\begin{figure}
    \centering
    \includegraphics[width=0.6\linewidth]{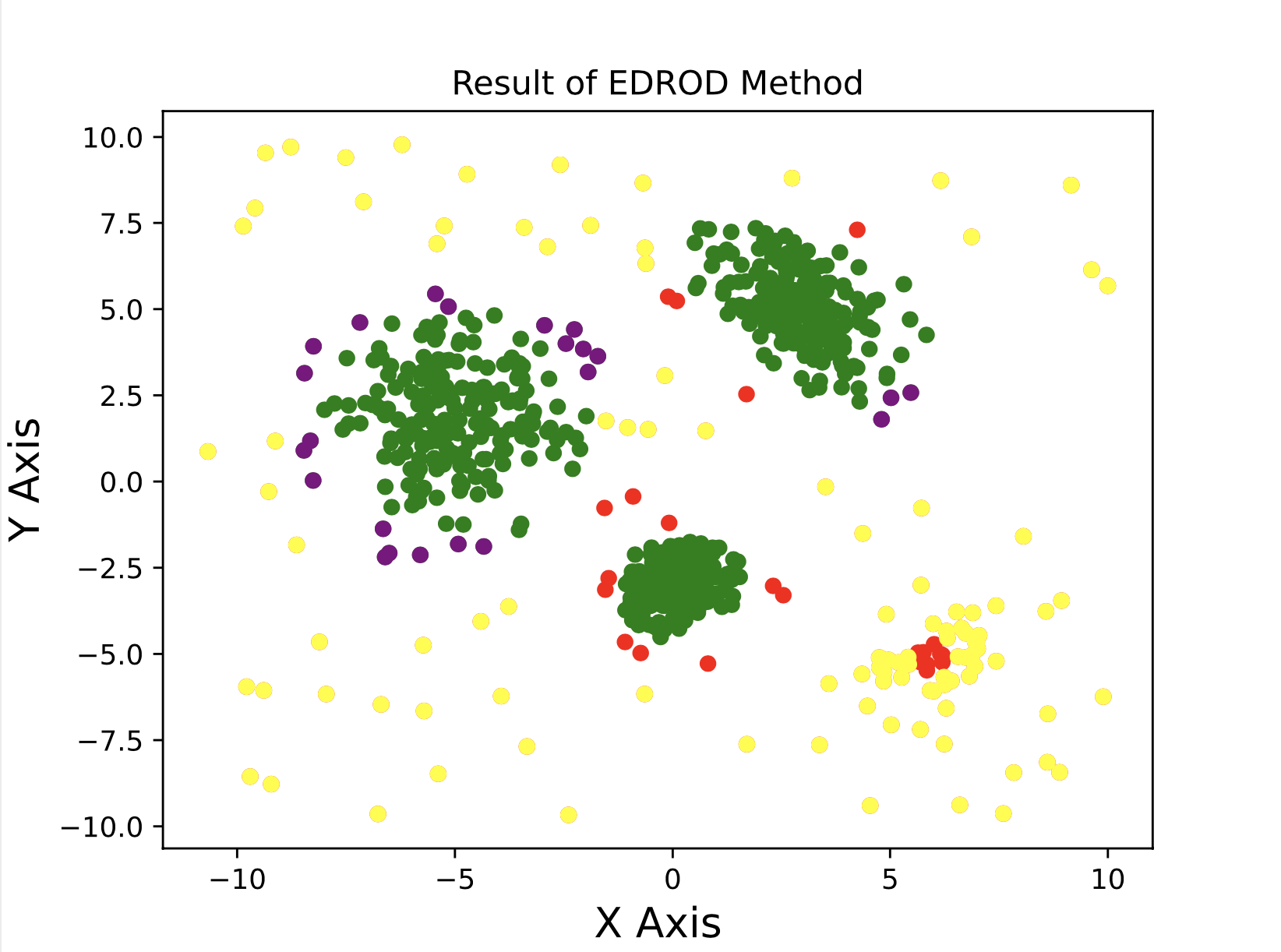}
    \caption{Visualization of detection results for the EDROD method on the two-dimensional dataset.}
    \label{fig:7}
\end{figure}

\begin{table}[ht]
\centering
\caption{Classification of samples using different colors}
\label{tab:3}
\begin{tabular}{c|c|c}
\hline
\textbf{Original sample category} & \textbf{Detection result} & \textbf{Color} \\
\hline
Normal & Normal & Green \\
\hline
Normal & Anomalous & Purple \\
\hline
Anomalous & Normal & Red \\
\hline
Anomalous & Anomalous & Yellow \\
\hline
\end{tabular}
\end{table}
According to TABLE \ref{tab:3}, the green color indicates that normal samples are correctly identified as normal, while the yellow color indicates that anomalous samples are correctly identified as anomalous. In addition, purple color means false positive identification for normal samples, while red color means false negative identification for anomalous samples.

In order to quantitatively compare the performance of various algorithms on this two-dimensional dataset, we calculated the number of samples for each color and recorded them in TABLE \ref{tab:4}.
\begin{table}[ht]
\centering
\caption{Number of samples for each color using different anomaly-detection methods}
\label{tab:4}
\begin{tabular}{c|cccc}
\hline
\textbf{Approach} & \textbf{Green} & \textbf{Red} & \textbf{Yellow} & \textbf{Purple} \\
\hline
KNN & 797 & 45 & 85 & 45 \\
LOF & 794 & 48 & 82 & 48 \\
ABOD & 801 & 41 & 89 & 41 \\
DCROD & 799 & 43 & 87 & 43 \\
OCSVM & 799 & 43 & 87 & 43 \\
COF & 788 & 54 & 76 & 54 \\
COPOD & 762 & 80 & 50 & 80 \\
PCA & 799 & 43 & 87 & 43 \\
ECOD & 818 & 24 & 106 & 24 \\
KDE & 799 & 43 & 87 & 43 \\
IForest & 803 & 39 & 91 & 39 \\
LUNAR & 788 & 54 & 56 & 54 \\
\textbf{EDROD} & \textbf{819} & \textbf{23} & \textbf{107} & \textbf{23} \\
\hline
\end{tabular}
\end{table}

According to the visual results as shown in Fig. \ref{fig_6}, and the quantitative results in TABLE \ref{tab:4}, all methods under study can be classified into four types according to the detection results for further analysis, which are Fig. \ref{fig_6}(a-e), (f-g), (h-i) and (j-l)\&Fig. \ref{fig:7} respectively. In the following, we shall analyze the detection results for each type of methods in detail.

The first type comprises algorithms that can only detect point anomalies but struggle to identify cluster anomalies, which include KNN, LOF, ABOD, DCROD and OCSVM (shown in Fig. \ref{fig_6} (a), (b), (c), (d) and (e) respectively). These algorithms incorporate KNN algorithms as one intermediate step to select the nearest neighbors for each data sample. Therefore, these algorithms primarily rely on local features within a local region to determine whether a sample is anomalous or not. Outliers usually exhibit distinct local features compared to cluster anomalies and normal samples. Consequently, these algorithms excel at detecting point anomalies.

The second type includes COF and COPOD algorithms, of which the detection results are shown in Fig. \ref{fig_6} (f) and (g) respectively. It can be observed that the two algorithms do not perform well in detecting both point anomalies and cluster anomalies. Taking the COF algorithm for example, the limitation of COF lies in the fact that in the process of detecting anomalies, COF evaluates the outlier score of each sample based on the information of neighborhood for this sample without considering the global context of the entire dataset. This may lead to inconsistencies when dealing with clusters or regions with different densities, making COF unable to identify cluster anomalies in the dataset.

The third type of algorithm, depicted in Fig. \ref{fig_6} (h) and (i), can effectively detect cluster anomalies but is less sensitive to point anomalies. This category includes PCA and ECOD methods. These methods employ the classification approach to determine anomalies based on classification results. Given the distinct characteristics of cluster anomalies, such as their appearance in the range $4.5 < x < 8$ and $-7 < y < -1$ in Fig. \ref{fig_6}, PCA and ECOD methods can successfully classify cluster anomalies. However, the two methods fail to detect many point anomalies.

Finally, the fourth type of algorithm, shown in Fig. \ref{fig_6}(j), (k), (l) and Fig. \ref{fig:7}, can effectively detect both point anomalies and cluster anomalies. This type of algorithm includes KDE, IForest, LUNAR and EDROD. Comparing results show that the EDROD method exhibits superior performance in terms of fewer misclassified samples, fewer missed samples and more correctly identified samples. EDROD only fails to detect nine cluster anomalies, while successfully detecting most point anomalies. Moreover, from TABLE \ref{tab:2}, we can find the EDROD method achieves the highest AUC value of 0.959. This is mainly because EDROD combines local entropy and global probability density, thus considering both local and global features to accurately detect outliers and cluster anomalies.

Next, we present the experimental results of the EDROD method and its competing counterparts on the ten-dimensional synthetic \textit{300\_SAMPLE} and \textit{700\_SAMPLE} datasets. We have to mention here that the value of kernel width in the Gaussian kernel function for kernel density estimation can affect the anomaly-detection performance of the EDROD method. In our studies, by using the grid search method, the optimal value of kernel width on the ten-dimensional synthetic datasets studied here is around 0.36. Therefore, we set the kernel width $h=0.36$ in the EDROD method for studying the synthetic datasets here.

\begin{figure}
    \centering
    \includegraphics[width=1\linewidth]{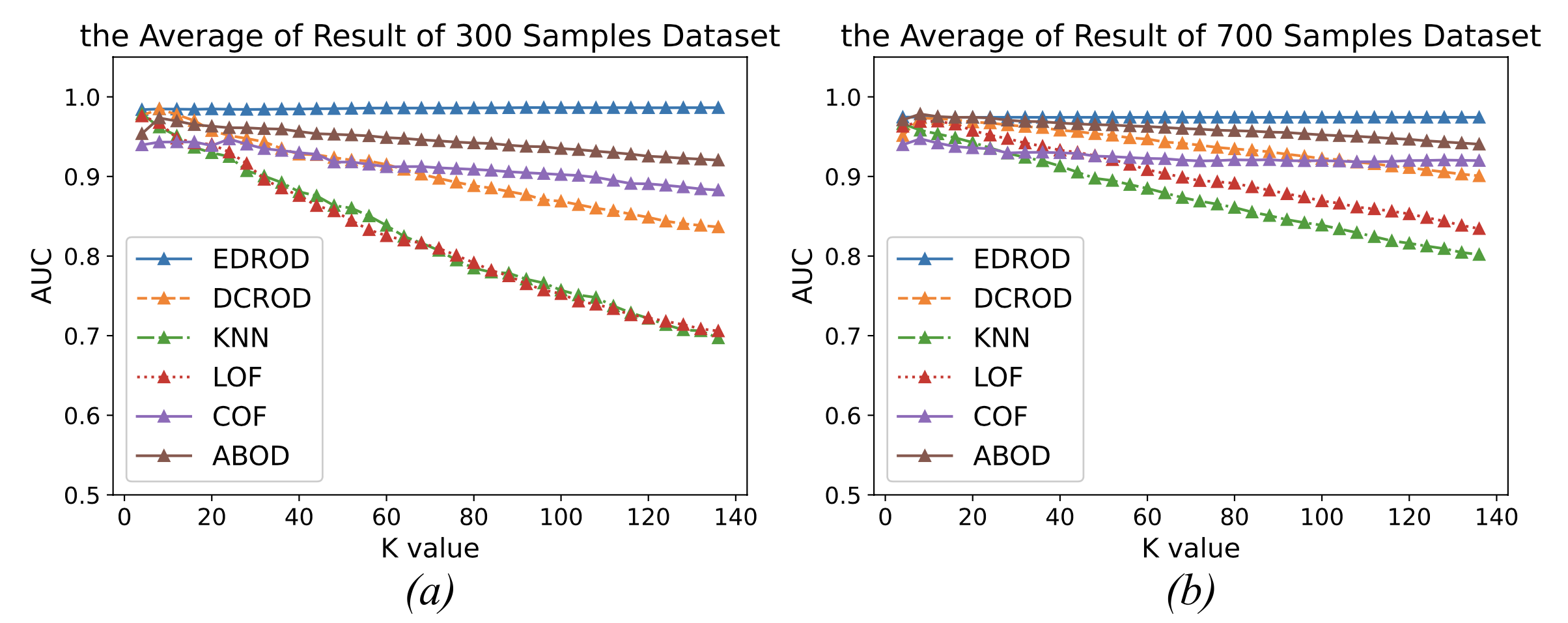}
    \caption{Average AUC values (over 10 different dataset instances) versus the parameter $K$ on the synthetic \textit{300\_SAMPLE} (a) and \textit{700\_SAMPLE} (b) respectively. }
    \label{fig:8}
\end{figure}

We first compare the performance of six KNN-based methods, which include DCROD, EDROD, KNN, LOF, COF, and ABOD. We vary the value of $K$ from 4 to 140. Fig. \ref{fig:8} shows the relationship between average AUC values and the parameter $K$ for all competing algorithms on these two synthetic datasets. It should be noted that each average AUC value is obtained over the results of all 10 independent dataset instances. It can be observed from both Fig. \ref{fig:8}(a) and Fig. \ref{fig:8}(b) that, as the value of $K$ changes, variations of resulting AUC values present different characteristics for all competing algorithms. KNN and LOF algorithms exhibit the most significant changes as the value of $K$ varies. Specifically, when $K$ is set to 140, the AUC value drops to below 0.7 (on the \textit{300\_SAMPLE} dataset) and around 0.8 (on the \textit{700\_SAMPLE} dataset) when using KNN and LOF algorithms. Thus, the effectiveness of both KNN and LOF algorithms is highly dependent on the value of $K$.
With a small sample size in the dataset, increasing the value of $K$ gradually leads to the estimation of both normal and anomalous samples together during the process of local kernel density estimation. As a result, the detection accuracy decreases, leading to a noticeable decline in AUC values. Besides, after comparing Fig. \ref{fig:8}(a) with Fig. \ref{fig:8}(b), we have also observed that the size of the dataset also has an impact on the accuracy of anomaly detection for LOF, KNN, and DCROD methods. It is evident that the AUC values of LOF, KNN, and DCROD algorithms on the \textit{300\_SAMPLE} dataset are lower than those on the \textit{700\_SAMPLE} dataset. This result indicates that the LOF, KNN, and DCROD algorithms are also sensitive to the change in dataset sizes. In contrast, EDROD exhibits high stability in terms of AUC values when varying the value of $K$. As $K$ increases from 4 to 140, the AUC values only fluctuate within an order of magnitude of $10^{-4}$. Moreover, EDROD consistently maintains higher AUC values than all other competing algorithms on both \textit{300\_SAMPLE} and \textit{700\_SAMPLE} datasets. The AUC values remain to be close to 1 when varying the value of $K$ on both \textit{300\_SAMPLE} and \textit{700\_SAMPLE} datasets, which indicates that the EDROD algorithms are also robust to the change of dataset sizes.


On the synthetic \textit{300\_SAMPLE} and \textit{700\_SAMPLE} datasets, we have also conducted anomaly detection using various non-KNN-based algorithms, which include OCSVM, COPOD, KDE, PCA, IForest, ECOD, and LUNAR. To compare the performance of both KNN-based and non-KNN-based methods, we demonstrate the AUC values of all competing algorithms in TABLE \ref{tab:5} (on the \textit{300\_SAMPLE} dataset) and TABLE \ref{tab:6} (on the \textit{700\_SAMPLE} dataset). It should be noted that each AUC value of the KNN-based algorithm in TABLE \ref{tab:5} and \ref{tab:6} is the average over the results with the parameter $K$ ranging from 4 to 140. It can be observed from TABLE \ref{tab:5} and TABLE \ref{tab:6} that the EDROD method can yield the highest average AUC values over all 10 dataset instances.

\begin{figure}
    \centering
    \includegraphics[width=1\linewidth]{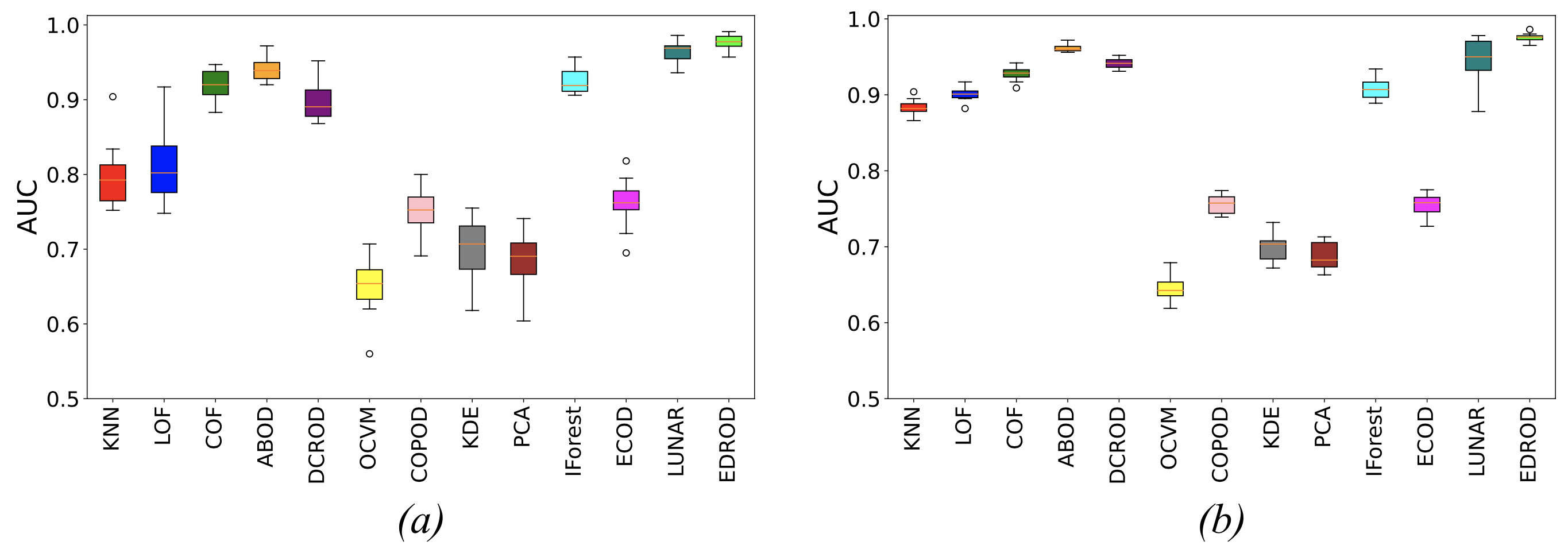}
    \caption{Boxplot of all KNN-based and non-KNN-based methods over all 10 dataset realizations for \textit{300\_SAMPLE} (a) and \textit{700\_SAMPLE} (b)  datasets respectively. }
    \label{fig:9}
\end{figure}

Then, based on TABLE \ref{tab:5} and TABLE \ref{tab:6}, we draw the boxplot of AUC values over the 10 datasets for all competing algorithms, which is demonstrated in Fig. \ref{fig:9}(a) (on the \textit{300\_SAMPLE} dataset) and Fig. \ref{fig:9}(b) (on the \textit{700\_SAMPLE} dataset) respectively. It can be found that the EDROD method can achieve the highest median AUC values. The algorithm that comes closest to the performance of the EDROD algorithm on the \textit{300\_SAMPLE} dataset is LUNAR. However, the LUNAR algorithm lags behind EDROD by 1.3\% on the \textit{300\_SAMPLE} in terms of median AUC values. In addition, the algorithm that comes closest to the performance of the EDROD algorithm on the \textit{700\_SAMPLE} dataset is ABOD, which lags behind EDROD by 1.5\% in terms of median AUC values. Besides, we have also found that the EDROD algorithm exhibits the smallest range of fluctuations compared to other algorithms, which means that the EDROD algorithm has the most stable performance when faced with datasets of different sizes. To sum up, for the synthetic \textit{300\_SAMPLE} and \textit{700\_SAMPLE} datasets, the EDROD method can achieve a higher AUC value and more stable performance of anomaly detection when compared to other competing algorithms.
\begin{table*}[!t]
\centering
\caption{AUC values of all competing algorithms on 10 different dataset instances of the \textit{300\_SAMPLE} dataset}
\label{tab:5}
\begin{tabular}{c|ccccccccccccc}
\hline
\multirow{2}{*}{Dataset} & \multirow{2}{*}{KNN} & \multirow{2}{*}{LOF} & \multirow{2}{*}{COF} & \multirow{2}{*}{ABOD} & \multirow{2}{*}{DCROD} & \multirow{2}{*}{OCSVM} & \multirow{2}{*}{COPOD} & \multirow{2}{*}{KDE} & \multirow{2}{*}{PCA} & \multirow{2}{*}{IForest} & \multirow{2}{*}{ECOD} & \multirow{2}{*}{LUNAR} & \multirow{2}{*}{EDROD} \\
 & & & & & & & & & & & & \\
\hline
1	&	0.812	&	0.810	&	0.911	&	0.944	&	0.897	&	0.690	&	0.782	&	0.739	&	0.718	&	0.915	&	0.755	&	0.972	&	\textbf{0.986} \\
2	&	0.809	&	0.842	&	0.940	&	0.946	&	0.915	&	0.660	&	0.735	&	0.727	&	0.692	&	0.911	&	0.795	&	0.936	&	\textbf{0.957} \\
3	&	0.834	&	0.859	&	0.947	&	0.965	&	0.935	&	0.707	&	0.800	&	0.755	&	0.741	&	0.957	&	0.818	&	0.986	&	\textbf{0.991} \\
4	&	0.760	&	0.748	&	0.929	&	0.934	&	0.883	&	0.560	&	0.691	&	0.618	&	0.604	&	0.906	&	0.695	&	0.971	&	\textbf{0.975} \\
5	&	0.904	&	0.917	&	0.942	&	0.972	&	0.952	&	0.676	&	0.761	&	0.732	&	0.713	&	0.931	&	0.766	&	0.972	&	\textbf{0.986} \\
6	&	0.776	&	0.772	&	0.899	&	0.929	&	0.876	&	0.636	&	0.713	&	0.687	&	0.662	&	0.940	&	0.721	&	0.973	&	\textbf{0.981} \\
7	&	0.813	&	0.826	&	0.931	&	0.951	&	0.907	&	0.662	&	0.766	&	0.728	&	0.689	&	0.942	&	0.781	&	0.952	&	\textbf{0.971} \\
8	&	0.776	&	0.794	&	0.909	&	0.921	&	0.871	&	0.620	&	0.736	&	0.674	&	0.653	&	0.912	&	0.752	&	0.967	&	\textbf{0.973} \\
9	&	0.761	&	0.784	&	0.906	&	0.928	&	0.884	&	0.632	&	0.744	&	0.673	&	0.679	&	0.923	&	0.758	&	0.963	&	\textbf{0.980} \\
10	&	0.752	&	0.773	&	0.883	&	0.920	&	0.868	&	0.648	&	0.771	&	0.666	&	0.694	&	0.907	&	0.769	&	0.947	&	\textbf{0.969} \\
\hline
Average	&	0.800	&	0.813	&	0.920	&	0.941	&	0.899	&	0.649	&	0.750	&	0.700	&	0.685	&	0.924	&	0.761	&	0.964	&	\textbf{0.977} \\
\hline
\end{tabular}
\end{table*}
\begin{table*}[!t]
\centering
\caption{AUC values of all competing algorithms on 10 different dataset instances of the \textit{700\_SAMPLE} dataset}
\label{tab:6}
\begin{tabular}{c|ccccccccccccc}
\hline
\multirow{2}{*}{Dataset} & \multirow{2}{*}{KNN} & \multirow{2}{*}{LOF} & \multirow{2}{*}{COF} & \multirow{2}{*}{ABOD} & \multirow{2}{*}{DCROD} & \multirow{2}{*}{OCSVM} & \multirow{2}{*}{COPOD} & \multirow{2}{*}{KDE} & \multirow{2}{*}{PCA} & \multirow{2}{*}{IForest} & \multirow{2}{*}{ECOD} & \multirow{2}{*}{LUNAR} & \multirow{2}{*}{EDROD} \\
 & & & & & & & & & & & & \\
\hline
1	&	0.871	&	0.897	&	0.925	&	0.958	&	0.937	&	0.624	&	0.739	&	0.680	&	0.663	&	0.894	&	0.758	&	0.940	&	\textbf{0.974} \\
2	&	0.878	&	0.895	&	0.923	&	0.958	&	0.936	&	0.635	&	0.750	&	0.683	&	0.677	&	0.895	&	0.734	&	0.972	&	\textbf{0.978} \\
3	&	0.866	&	0.882	&	0.917	&	0.956	&	0.931	&	0.619	&	0.740	&	0.672	&	0.673	&	0.889	&	0.727	&	0.933	&	\textbf{0.965} \\
4	&	0.879	&	0.896	&	0.909	&	0.958	&	0.936	&	0.679	&	0.765	&	0.725	&	0.709	&	0.917	&	0.775	&	0.966	&	\textbf{0.976} \\
5	&	0.904	&	0.917	&	0.942	&	0.972	&	0.952	&	0.676	&	0.761	&	0.732	&	0.713	&	0.931	&	0.766	&	0.972	&	\textbf{0.986} \\
6	&	0.882	&	0.906	&	0.933	&	0.960	&	0.942	&	0.646	&	0.770	&	0.700	&	0.688	&	0.916	&	0.773	&	0.960	&	\textbf{0.972} \\
7	&	0.882	&	0.903	&	0.933	&	0.960	&	0.941	&	0.637	&	0.742	&	0.687	&	0.673	&	0.903	&	0.745	&	0.932	&	\textbf{0.977} \\
8	&	0.886	&	0.905	&	0.929	&	0.965	&	0.947	&	0.654	&	0.766	&	0.708	&	0.709	&	0.934	&	0.762	&	0.978	&	\textbf{0.980} \\
9	&	0.889	&	0.899	&	0.928	&	0.959	&	0.944	&	0.652	&	0.774	&	0.707	&	0.695	&	0.902	&	0.758	&	0.928	&	\textbf{0.971} \\
10	&	0.895	&	0.905	&	0.936	&	0.965	&	0.947	&	0.639	&	0.754	&	0.707	&	0.675	&	0.911	&	0.749	&	0.878	&	\textbf{0.976} \\
\hline
Average	&	0.883	&	0.901	&	0.928	&	0.961	&	0.941	&	0.646	&	0.756	&	0.700	&	0.688	&	0.909	&	0.755	&	0.946	&	\textbf{0.976} \\
\hline
\end{tabular}
\end{table*}

Based on the experiments conducted on synthetic datasets, the following conclusions can be drawn. The proposed EDROD method shows lower sensitivity to the choice of parameter $K$ compared to other KNN-based algorithms, indicating enhanced robustness to the change of parameter $K$. Moreover, the detection accuracy of EDROD is hardly affected by the size of the dataset. Finally, when compared to non-KNN-based algorithms, EDROD can also achieve higher AUC values, which indicates that EDROD can achieve more accurate detection for anomalies on the synthetic dataset we study.

\subsection{Experimental results and discussions on the real-world datasets}
Due to the prevalence of numerous zero values in real-world datasets, the ABOD algorithm is not applicable and therefore not included in the comparative evaluation with other methods on the real-world datasets. In addition, since data distributions on different datasets vary from each other, careful consideration must be given to assigning appropriate values to the kernel width in the Gaussian kernel function for kernel density estimation. In our experiments, we utilize the grid optimization method to find the optimal value of kernel width for each dataset. The optimal values of kernel width employed in the nine real-world datasets are demonstrated in TABLE \ref{tab:7}.

\begin{table}[!ht]
\centering
\caption{Value of kernel width employed in the EDROD method on each real-world dataset}
\label{tab:7}
\begin{tabular}{l c | l c}
\hline
\textbf{Dataset} & \textbf{Kernel width} & \textbf{Dataset} & \textbf{Kernel width} \\
\hline
Wind turbine & 1.25 & Letter & 3.5 \\
Wave & 3.7 & Ionosphere & 0.75 \\
Breastw & 5 & Parkinson & 0.55 \\
WDBC & 29 & Glass & 0.56 \\
Wine & 85 & & \\
\hline
\end{tabular}
\end{table}

\begin{figure*}[!t]
\centering
\includegraphics[width=1\linewidth]{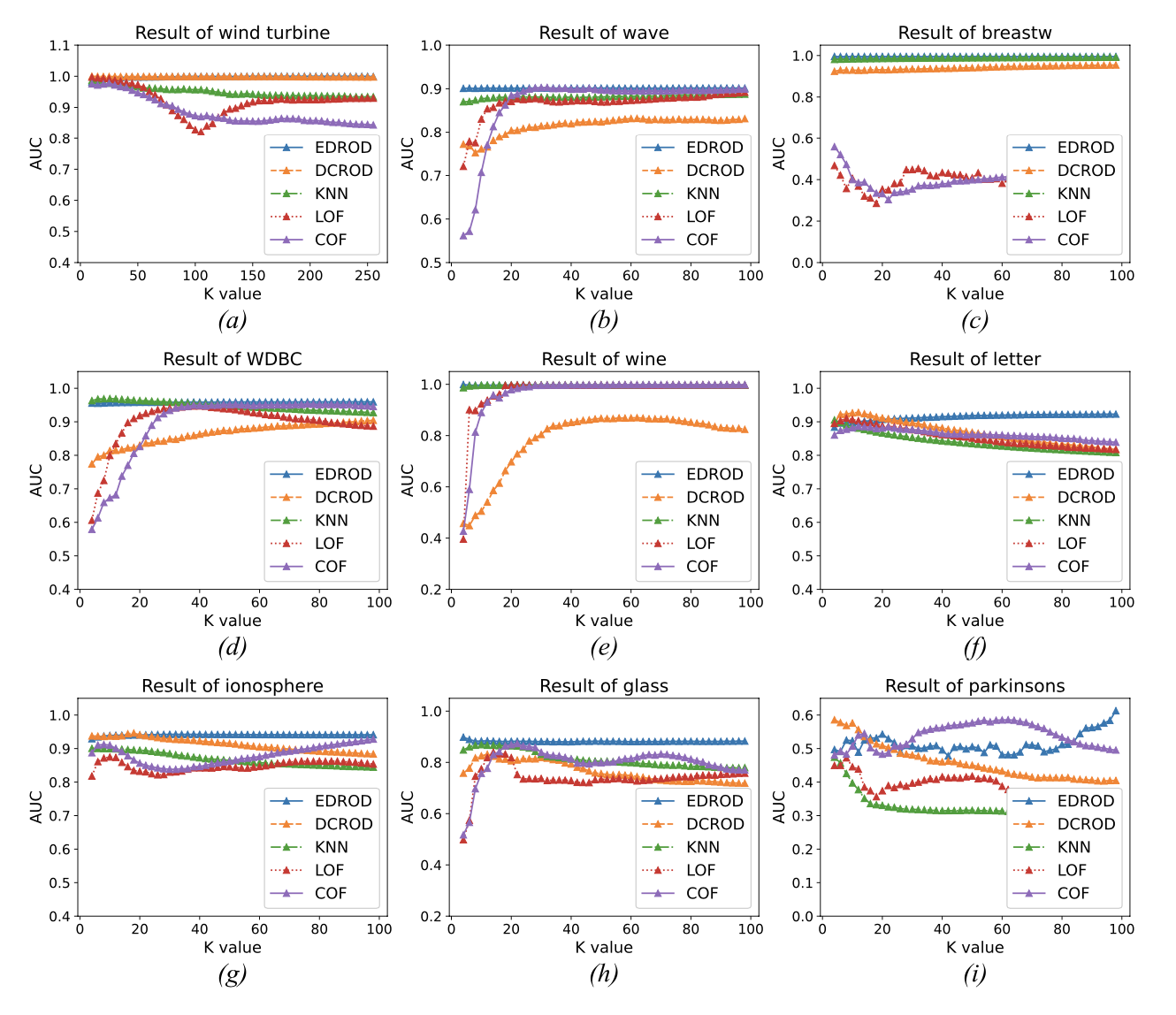}
\caption{AUC values versus the parameter K on nine real-world datasets}
\label{fig_10}
\end{figure*}

Since all methods present performance differences in a wider range of parameter K on the wind turbine dataset, the value of $K$ ranges from 20 to 260 for this particular dataset. For the remaining datasets, the value of $K$ ranges from 4 to 100. The variations of AUC values with respect to $K$ for all KNN-based algorithms are presented in Fig. \ref{fig_10}. It can be observed that, except for the Parkinson dataset, the AUC values of the EDROD method exhibit strong robustness to changes of $K$ on other datasets. On datasets such as breastw, WDBC, wine, and glass, the AUC values of EDROD and KNN are close to each other when the value of $K$ is small. However, as $K$ increases, the AUC values of the KNN algorithm show significant fluctuations. This behavior can be attributed to the fact that when using only the KNN algorithm for anomaly detection, the decision criterion is solely based on local features. When $K$ is small, local features can be effective for detecting anomalies. However, as $K$ increases, the selected local features may differ from the true local features, resulting in the influence of normal samples on the detection of anomalous samples. In contrast, EDROD utilizes both local and global features as decision criteria, providing stronger robustness compared to the KNN algorithm. On the wave, WDBC, and ionosphere datasets, as the value of $K$ increases, the AUC values of the COF algorithm gradually approach those of EDROD. This is because as $K$ increases, the COF algorithm can leverage more sample features to detect anomalies. In contrast, EDROD incorporates global density probability values into its detection process, thus eliminating the need for increasing the value of $K$ to obtain global sample features. Finally, we have to mention that on the Parkinson dataset, the presence of a high proportion of anomalous samples affects the detection performance of all KNN-based algorithms. Consequently, the overall detection performances of these algorithms on the Parkinson dataset are not as good as those on other datasets. However, when compared to other KNN-based algorithms, EDROD exhibits more accurate anomaly-detection results and stronger robustness to the change of parameter $K$ on the Parkinson dataset.

Besides, we have also conducted comprehensive studies on these real-world datasets for non-KNN-based algorithms. AUC values of all KNN-based and non-KNN-based algorithms are demonstrated in TABLE \ref{tab:8}.  For each dataset, all algorithms are ranked in descending order based on their AUC values.
Over the total nine real-world datasets, EDROD achieves the impressive lowest average ranking of 1.41, located in the top position in all twelve methods, which signifies the exceptional detection accuracy of EDROD methods.

\begin{table*}[!t]
\centering
\caption{AUC values of all KNN-based and non-KNN-based methods on nine real-world datasets}
\label{tab:8}
\begin{tabular}{c|ccccccccccccc}
\hline
\multirow{2}{*}{Dataset} & \multirow{2}{*}{KNN} & \multirow{2}{*}{LOF} & \multirow{2}{*}{COF}  & \multirow{2}{*}{DCROD} & \multirow{2}{*}{OCSVM} & \multirow{2}{*}{COPOD} & \multirow{2}{*}{KDE} & \multirow{2}{*}{PCA} & \multirow{2}{*}{IForest} & \multirow{2}{*}{ECOD} & \multirow{2}{*}{LUNAR} & \multirow{2}{*}{EDROD} \\
 & & & & & & & & & & & & \\
\hline
wind turbine & 0.947 & 0.922 & 0.877 & 0.998 & 0.934 & \textbf{0.999} & 0.964 & 0.995 & 0.991 & 0.998 & 0.997 & \textbf{0.999} \\
wave & 0.882 & 0.871 & 0.875 & 0.836 & 0.879 & 0.860 & 0.879 & 0.837 & 0.836 & 0.784 & 0.764 & \textbf{0.901} \\
breastw & 0.988 & 0.433 & 0.399 & 0.929 & 0.979 & 0.994 & 0.976 & 0.959 & 0.985 & 0.991 & 0.970 & \textbf{0.995} \\
WDBC & 0.949 & 0.909 & 0.904 & 0.878 & 0.509 & 0.949 & 0.576 & 0.901 & 0.901 & 0.852 & 0.829 & \textbf{0.959} \\
wine & 0.997 & 0.989 & 0.984 & 0.789 & 0.475 & 0.867 & 0.756 & 0.821 & 0.778 & 0.733 & 0.469 & \textbf{0.997} \\
letter & 0.840 & 0.856 & 0.864 & 0.871 & 0.890 & 0.560 & 0.914 & 0.523 & 0.622 & 0.572 & 0.901 & \textbf{0.915} \\
ionosphere & 0.869 & 0.850 & 0.878 & 0.914 & 0.762 & 0.799 & 0.923 & 0.795 & 0.847 & 0.735 & 0.927 & \textbf{0.941} \\
Parkinson & 0.325 & 0.376 & 0.540 & 0.460 & 0.462 & 0.543 & 0.489 & 0.375 & 0.474 & 0.398 & \textbf{0.655} & 0.516 \\
glass & 0.815 & 0.751 & 0.813 & 0.770 & 0.593 & 0.645 & 0.817 & 0.603 & 0.725 & 0.621 & 0.827 & \textbf{0.882} \\
\hline
Rank & 5.444 & 7.333 & 6.333 & 6.667 & 8.444 & 5.444 & 5.778 & 8.778 & 7.556 & 8.667 & 6.073 & \textbf{1.410} \\
\hline
\end{tabular}
\end{table*}

\begin{figure}
    \centering
    \includegraphics[width=0.7\linewidth]{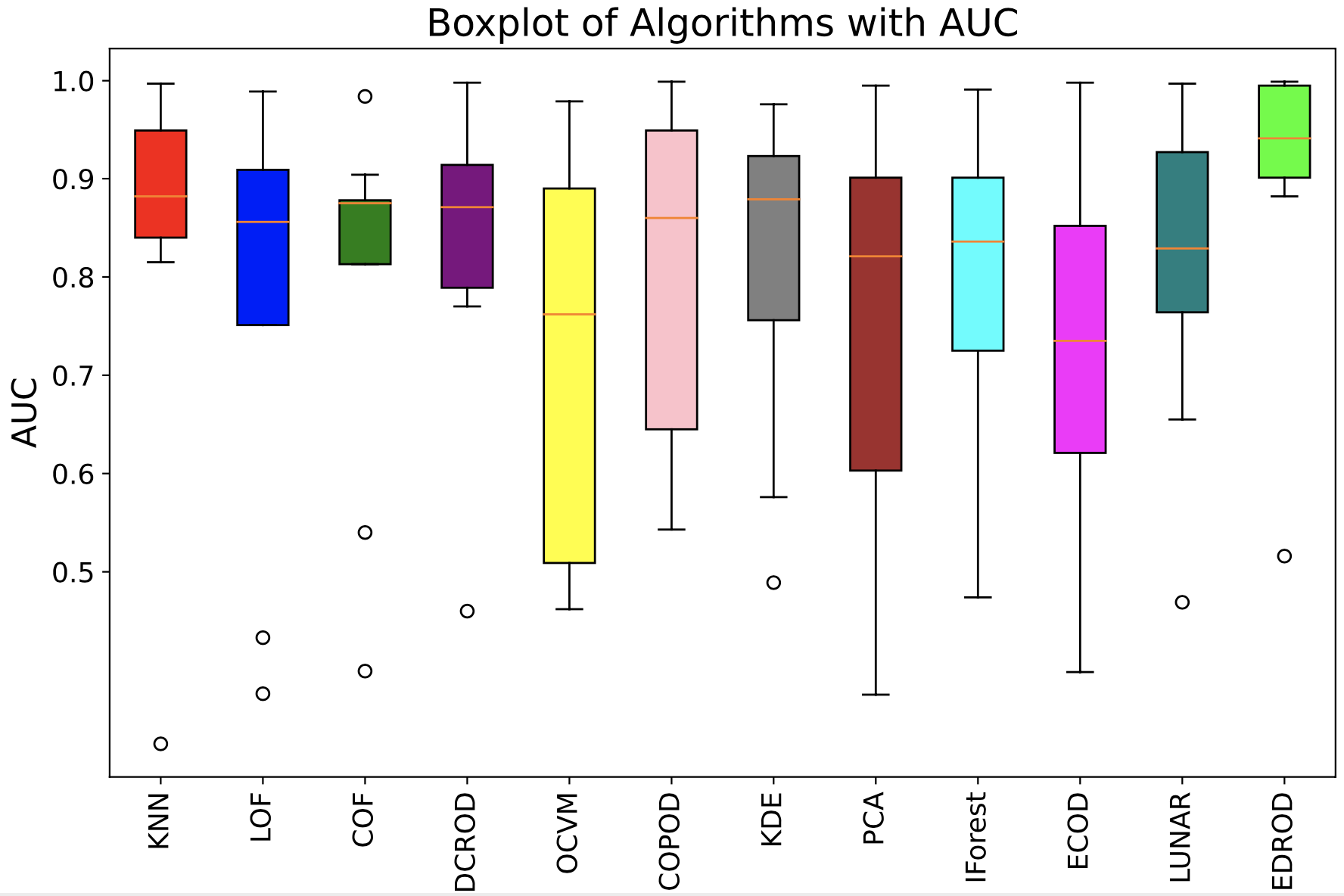}
    \caption{Boxplot of all KNN-based and non-KNN based methods over all nine real-world datasets.}
    \label{fig:11}
\end{figure}

To illustrate the AUC values shown in TABLE \ref{tab:8} in more intuitive way, we draw the boxplot of all KNN-based and non-KNN-based methods over all nine real-world datasets, which is shown in Fig. \ref{fig:11}. It can be clearly observed that our proposed EDROD method have the minimal median AUC values, which indicates again that the EDROD method can achieve the most accurate performance of anomaly detection on the nine real-world datasets. In addition, we have also found from the boxplot that the EDROD method results in small box, indicating that the EDROD method can be applied to detect anomalies on a wide variety of datasets. However, the Parkinson dataset is the exception since more than 75\% of all samples on the Parkinson dataset are anomalies. All methods studied here, including our proposed EDROD method, do not achieve high detection accuracy for anomaly detection on the Parkinson dataset.

\section{Conclusion}
In this paper, we propose a novel robust method for anomaly detection, called Entropy Density Ratio Outlier Detection (EDROD) for short. By combining both global and local features of data samples in the dataset, the EDROD method is capable of simultaneously identifying point anomalies and cluster anomalies. The global feature of each sample is represented by the global density which is obtained by using KDE method. The local feature of each sample are represented by the local Shannon entropy which is calculated based on densities of this sample and its \textit{K} nearest neighboring samples. By integrating global and local features, the \textit{EDR} (Entropy Density Ratio) score is computed, indicating the degree of anomaly of each sample within the dataset. A higher \textit{EDR} score indicates a higher likelihood of being an anomalous sample.

The high effectiveness of the EDROD method is verified by comparing EDROD with several other anomaly detection algorithms on both synthetic datasets and real-world datasets. Experimental results demonstrate the high accuracy and enhanced robustness of the EDROD method across diverse dataset sizes, data distributions, and ratios of anomalous samples in the dataset. Regarding the sensitivity of KNN-based algorithms to the selection for the value of $K$ (i.e. the number of nearest neighboring samples), this paper specifically analyzes the impact of the parameter $K$ on the performance of the EDROD method. By comparing the performance of the EDROD method with those of some other KNN-based algorithms, it is found that the EDROD method exhibits enhanced robustness to value variations in $K$.

Although the proposed EDROD method successfully solves the problem of simultaneously detecting point anomalies and cluster anomalies, there is still much room to further improve the effectiveness of anomaly-detection algorithms. For example, all methods mentioned in this paper, including our proposed EDROD method, cannot efficiently detect anomalies on datasets with a high proportion of anomalous samples like the Parkinson dataset. Moreover, EDROD exhibits a relatively slow detection speed on large datasets. Future research can focus on improving the detection speed and enhancing the capability to handle datasets with a high proportion of anomalous samples.


\section*{Acknowledgments}
This work was supported by the National Key R\&D Program of China (2022YFE0198900), the National Natural Science Foundation of China (61771430, 62176236), and the Natural Science Foundation of Zhejiang Province under grant LY22F020015.

\vfill

\end{document}